\documentclass[conference]{IEEEtran}
\IEEEoverridecommandlockouts
\usepackage{cite}
\usepackage{amsmath,amssymb,amsfonts}
\usepackage{algorithmic}
\usepackage{multirow}
\usepackage{graphicx}
\usepackage{textcomp}
\usepackage{xcolor}
\def\BibTeX{{\rm B\kern-.05em{\sc i\kern-.025em b}\kern-.08em
    T\kern-.1667em\lower.7ex\hbox{E}\kern-.125emX}}
\begin{document}

\title{Matlab-based Epoch Extraction for Speaker Differentiation\\
}
\author{\IEEEauthorblockN{ Kunlun Li}
\IEEEauthorblockA{\textit{Department of Electrical and Computer Engineering} \\
\textit{The University of Mississippi}\\
Oxford, USA}
\and
\IEEEauthorblockN{Daniel Ferro}
\IEEEauthorblockA{\textit{Department of Electrical and Computer Engineering} \\
\textit{The University of Mississippi}\\
Oxford, USA}
\and
\IEEEauthorblockN{Xu Zhao}
\IEEEauthorblockA{\textit{Department of Electrical and Computer Engineering} \\
\textit{The University of Mississippi}\\
Oxford, USA}
\and
\IEEEauthorblockN{Abdul Jabbar Syed}
\IEEEauthorblockA{\textit{LTRC} \\
\textit{International Institute of Information Technology Hyderabad}\\
Hyderabad, India}
\and
\IEEEauthorblockN{Anil K Vuppala}
\IEEEauthorblockA{\textit{LTRC} \\
\textit{International Institute of Information Technology Hyderabad}\\
Hyderabad, India}
\and
\IEEEauthorblockN{Azeemuddin Syed}
\IEEEauthorblockA{\textit{Centre for VLSI and Embedded Systems} \\
\textit{International Institute of Information Technology Hyderabad}\\
Dept. of ECE, The University of Mississippi \\
syed@iiit.ac.in, asyed@olemiss.edu
}
}

\maketitle

\begin{abstract}
Epoch extraction has become increasingly popular in recent years for speech analysis research because accurately detecting the location of the Epoch is crucial for analyzing speech signals. The Epoch, occurring at the instant of excitation in the vocal tract system, particularly during glottal closure, plays a significant role in differentiating speakers in multi-speaker conversations. However, the extraction of the Epoch poses a challenge due to the time-varying factors in the vocal tract system, which makes deconvolution for obtaining the original excitation location more complex. In this paper, various methods for Epoch extraction, including Zero Frequency Filtering (ZFF) and Zero Frequency Resonator (ZFR), will be discussed, and their pros and cons evaluated. In addition, the stability, accuracy, and feasibility of each method will be compared. The evaluation will involve a Matlab-based locking algorithm, and a proposed hardware implementation using Raspberry pi for speaker differentiation. The experiment includes six individuals uttering the phrase "The University of Mississippi," with one person acting as the reference or "lock" speaker. The number of epochs occurring at similar positions to the reference speaker will be counted as Delta, with larger Delta values indicating greater speaker similarity. Experimental results demonstrate that when the speaker remains the same, the average number of Delta is 7.5, while for different speakers, the average number of Delta decreases to 3, 2, 2, 2, and 1, respectively, representing a decrease of approximately 73\% in the number of epochs at similar positions compared to the reference speaker.

\end{abstract}


\section{Introduction}
  The past few decades have witnessed the continuous pursuit of accurate speaker differentiation in speech. Detection of the Epoch location in the speech is an essential part of speaker verification and differentiation. The instant of significant excitation of the vocal-tract system is referred to as the epoch\cite{b1}. The excitation means the magnitude of the signal at a specific point is substantially larger than the nearby points. In the speech, such excitation always takes place at the glottal closure instant (GCI). GCI is the instant when the glottis in the vocal tract system is stopped. On that occasion, the airflow is obstructed to pass the vocal tract. Since there is no airflow across the glottis, the airflow will cross the vocal fold and cause vibration. Hence the impulse-like signal will be generated which is referred to as excitation in the speech signal.
  
  Determining the epoch locations from speech signals is helpful in glottal source analysis\cite{b2}, glottal inverse filtering\cite{b3}, speech analysis, speaker verification and differentiation, speech synthesis, and pitch perception. There are some state-of-the-art methods to extract the Epoch such as Linear prediction (LP), and dynamic programming projected phase-slope algorithm (DYPSA)\cite{b4}. The basis of these methods is that LP residuals highly contain the excitation of the original speech. These advanced techniques have robust performance in the detection of GCI. However, the complexity of such methods is high, and they also highly depend on the ability to model the system.
  
  There are also some traditional methods for epoch detection and extraction. Among these methods, Zero Frequency Resonator (ZFR)\cite{b1} and Zero Frequency Filtering (ZFF)\cite{b1} are the simplest and most widely used methods to detect and extract epochs with accuracy and efficiency. The basis for these ideas is filtering the speech signal at a specific frequency, in other words, passing the speech signal to a special narrow bandwidth filter which is at a specific frequency. Since filtering at zero frequency will not affect the character of the discontinuities and it can also reduce the impact of resonances at high frequency, Zero Frequency Filtering has been shown to be robust in the estimation of epochs location\cite{b5}.
  
  Compared with the Zero Frequency Filtering, Zero Frequency Resonator is easier to design and be implemented. But the non-linear phase characteristic of the method will cause the phase distortion of the output, which will impact the location of the detected epoch. However, since the IIR implementation of the Zero Frequency Filtering method, the system is not stable. Hence, Zero-Phase Zero Frequency Resonator (ZP-ZFR)\cite{b1} method is proposed. ZP-ZFR combines the advantages of both ZFF and ZFR. The ZP-ZFR has a linear phase and stable system. Hence the ZP-ZFR will guarantee stability without phase distortion. But the restriction of the zero phase increases the complexity of the method. 
  
  To evaluate the function of the above methods, the CMU Arctic Database will be used to verify the results. That database provided an audio signal as well as a complimentary electroglottograph (EGG) signal. The negative peaks of the EGG can be calculated to find the location of the epochs. The functionality of the above methods can be evaluated by comparing whether the detected location of epochs is the same as the result derived from EGG.
  
  In order to use the detected epoch to do the speaker differentiation, the lock algorithm will also be proposed via Matlab to implement the above method into the Raspberry pi.  By comparing the number of epochs at similar positions from different speakers, the speaker will be identified by comparing the number with a defined threshold. 
  
  The structure of the paper is organized as follows: Section II provides a comprehensive illustration of the basis of the proposed method. Section III will elaborate on and compare the ZFR, ZFF, and ZP-ZFR methods. In Section IV, the locking algorithm will be proposed and the Raspberry pi and Matlab will be implemented to use epoch location to differentiate the speakers. The conclusion and discussion will be provided in Section V.

\section{Extract at zero frequency}

\subsection{Challenges in Epoch detection}
Most of the epochs in the speech signal are produced by the vibration of the glottis. Most of the excitation takes place near the instance when the glottis is vibrating. At that moment, a large energy will be released which can be regarded as an impulse-like signal. Hence, the excitation can be regarded as an impulse signal in the time domain.

However, even if the magnitude and the time are explicit after treating the excitation like an impulse-like signal as shown in Fig. 1. The characteristic of the impulse signal is not clear anymore after modulated by the vocal tract system during the time-varying feature. As shown in Fig.1, before modulating by the vocal tract system H(z), the time of each impulse is clear, which can be represented as $t_1$, $t_2$, $t_3$…$t_k$. However, after passing the original excitation into a vocal tract system H(z), the location of each impulse is not explicit anymore. Therefore, one of the challenges in Epoch detection is that the speech signal analysis is a blind deconvolution if the aim is to find original excitation. Because the output signal is derived by conducting convolution between the original excitation and vocal tract system. But in this case, neither the original signal nor the impulse response is known.
\begin{figure}[htbp]
\centerline{\includegraphics[width=0.5\textwidth]{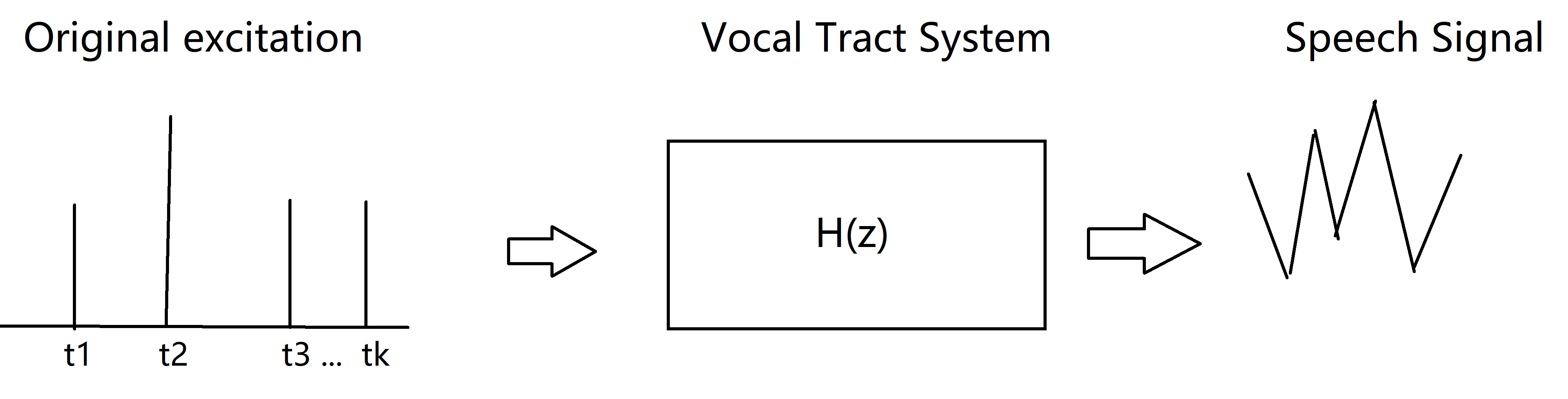}}
\caption{Transition of excitation to speech signal}
\label{fig}
\end{figure}

\subsection{Considered solution to the challenge}
Some basis and theories should be considered to solve that challenge. First, the excitation can be approximated as an impulse-like signal in the time domain. When using Fourier Transform to transform the impulse signal from the time domain to the frequency domain, the transformed impulse signal will spread across all frequencies in the frequency domain as shown in Fig.2. Second, the excitation in the time domain means that there is discontinuity in the time domain. Because the excitation can be interpreted as the magnitude at a specific point is substantially larger than the neighborhood. Hence the time domain will contain the characteristic of discontinuity after being excited by impulse-like excitation.
\begin{figure}[htbp]
\centerline{\includegraphics[width=0.5\textwidth]{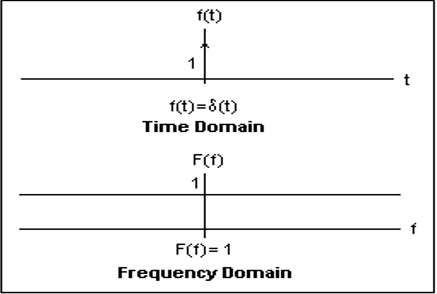}}
\caption{Impulse signal in time and frequency domain}
\label{fig2}
\end{figure}

The relationship between the signal in the time domain and the frequency domain is shown in Fig.3. After dividing the signal in the time domain into plenty of different components, each component will correspond to a specific frequency in the frequency domain. Hence, each specific frequency will remain characteristic of discontinuity in the time domain. To highlight the feature of discontinuities, the special filter should be designed with an extreme narrow bandwidth so that it can only pass the specific frequency and filter  all the other irrelevant information.
\begin{figure}[htbp]
\centerline{\includegraphics[width=0.5\textwidth]{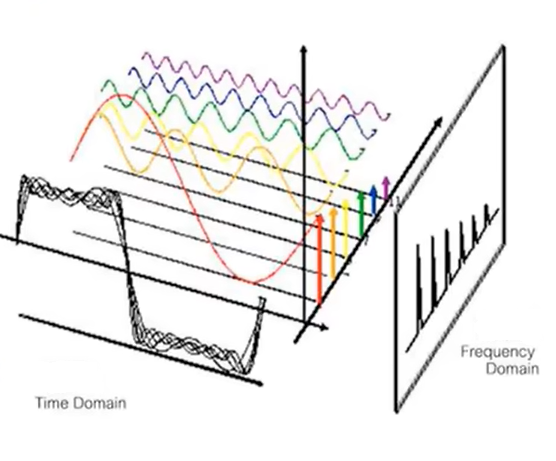}}
\caption{Relationship between time and frequency domain}
\label{fig3}
\end{figure}

\subsection{Extract at zero frequency}
The information on discontinuity will equally spread in the frequency domain due to the characteristic of the impulse signal, so every frequency will include the information on discontinuity including 0Hz. The advantage of choosing a zero-frequency resonator is that the characteristics of the time-varying vocal tract system will not affect the characteristic of the discontinuities in the resonator filter output\cite{b1}. Therefore, the proposed method to highlight the information of discontinuity is to extract the information at 0Hz because it can reduce the effect of resonance at high frequency. 

\section{Proposed Methods}
Based on the considered solution above, some methods regarding filtering speech signals at 0Hz are proposed.  Zero Frequency Resonator (ZFR), Zero Frequency Filtering (ZFF), and Zero Phase-Zero Frequency Resonator (ZP-ZFR) are three proposed methods to detect the GCI in the speech signal. These methods are robust because they can filter other irrelevant information and only preserve the energy around zero frequency. The information about discontinuity excited by the impulse-like signal is also included by the zero frequency. After getting filtered output, the positive zero crossing in the output signal indicates the location of the epochs because the slope of the signal is the largest at the positive zero crossing, which means the filtered signal has the fastest-changing speed at the positive zero crossing point. In addition, considering the feature of the impulse-like signal or the excitation, these signals indicate releasing high energy at a specific instant, which can be regarded as the fastest change as well.  Hence, these methods can effectively extract and locate the position of the epochs.

\subsection{Proposed Zero Frequency Resonator}\label{AA}
Zero Frequency Resonator (ZFR) is the first proposed method depending on extracting information around zero frequency. As mentioned above, the impulse-like signal of the excitation in the time domain will spread equally in the frequency domain. In addition, the excitation in the time domain could also be regarded as discontinuities because of the characteristic of excitation. Hence every single frequency in the frequency domain including zero frequency will contain the information of discontinuities. Therefore, using Zero Frequency Resonator can manifest such discontinuities, and based on the positive zero-crossing in the result, the location of epochs can be detected as well. 
\begin{itemize}
\item The equation of ZFR is given as
\begin{equation}
y[n]=-\sum_{k=1}^{2} a_ky[n-k]+x[n]\label{eq1}   
\end{equation}
In Eq.1, the $a_1$=-2 and $a_2$=1.
Plug the $a_1$ and $a_2$ into the equation, the result is shown as
\begin{equation}
y[n]=x[n]+2y[n-1]-y[n-2]\label{eq2}   
\end{equation}
So the data flow graph of zero frequency resonator is drawn as
\begin{figure}[htbp]
\centerline{\includegraphics[width=0.5\textwidth]{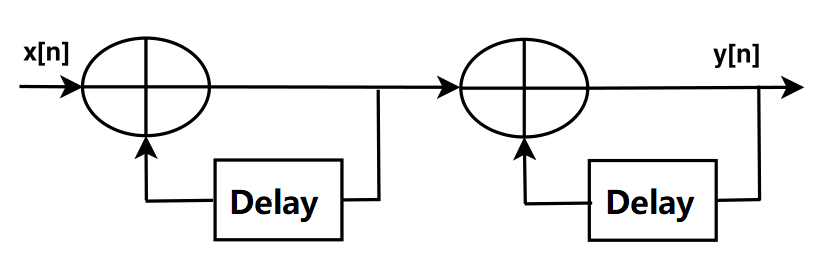}}
\caption{Data flow graph of zero frequency resonator}
\label{fig4}
\end{figure}
\item Since the signal should be analyzed in the discrete-time domain, it is important to transfer the equation from the time domain into the z domain by using the z transform shown in Eq.3.
\begin{equation}
X(Z)=\sum_{n=-\infty}^{\infty}x[n]Z^{-n}\label{eq3}   
\end{equation}
The Eq.3 shows the procedure to transform the signal from time domain to z domain, after apply the equation, the result of Eq.2 should be 
\begin{equation}
y_{ZFR}(z)=\frac{1}{1-2z^{-1}-z^{-2}}\label{eq4}   
\end{equation}
\item In the proposed ZFR method, there are two ZFR in cascade since a cascade of two 0-Hz resonators provide a sharper roll-off compared to a single 0-Hz resonator\cite{b1}. Hence, the resulting function of the ZFR method is shown as
\begin{equation}
H_{ZFR}(z)=\frac{1}{(1-rz^{-1})^4}\label{eq5}   
\end{equation}
In the Eq.5, the r is introduced as a constant selected from 0 to 1. Because the r determines the bandwidth of the system, when the r decreases to 0, the transfer function will become 1 and the system will lose the function of the filter. Hence, it is desirable to choose the value of r as higher as possible. In the experiment, it is ideal to select r between 0.95 to 0.99. For a value lower than 0.90, the false alarms will increase as the value of r declines.     
\item The advantage of the proposed ZFR method is that the structure is simple because there are only two ZFRs in the system. However, one of the drawbacks of that method is inaccurate because of phase distortion.

To make sure the accuracy of the result, the phase distortion should be eliminated. Hence the filter or the whole system should be linear-phase. In the method proposed above, the function of ZFR in the frequency domain is given as
\begin{equation}
H_{ZFR}(e^{j\omega})=\frac{1}{(1-rcos(\omega)+jrsin(\omega))^4}\label{eq6}   
\end{equation}
And the magnitude and phase of the system are given as 
\begin{equation}
|H_{ZFR}(e^{j\omega})|=\frac{1}{((1-rcos(\omega))^2+(rsin(\omega))^2)^2}\label{eq7}   
\end{equation}
And the magnitude and phase of the system are given as 
\begin{equation}
\theta(\omega)=-4tan^{-1}(\frac{rsin(\omega)}{1-rcos(\omega)})\label{eq8}   
\end{equation}
Based on Eq.8, the phase plot of the ZFR method is shown in Fig.5
\begin{figure}[htbp]
\centerline{\includegraphics[width=0.5\textwidth]{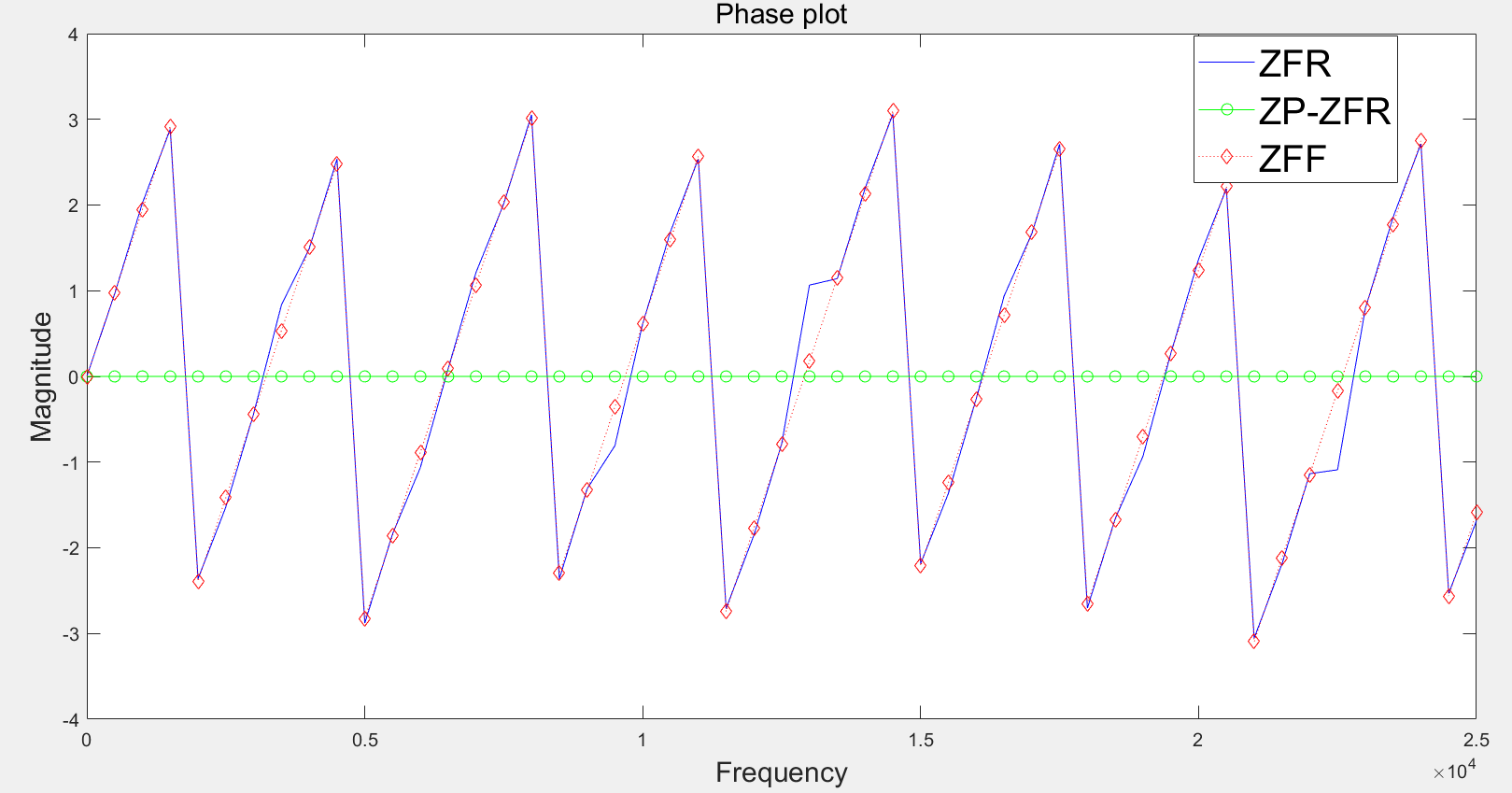}}
\caption{Phase plot for different methods}
\label{fig5}
\end{figure}
The phase plot shown in Fig.5 shows the phase of the ZFR is not a linear phase, which means the phase distortion will occur during the extraction of the epochs. Therefore, the location of the epoch will be varied due to the distortion. Since epoch extraction needs accuracy of the location of the epoch, the ZFR is not the best way to extract the location of the epoch due to the characteristic of non-linear. To extract epochs accurately, another method with linear phase should be proposed which is Zero Frequency Filtering (ZFF) method.
\end{itemize}

\subsection{Consideration of Zero Filtering method}
Zero Frequency Filtering (ZFF) is one of the most accurate and efficient ways nowadays to detect and extract the location of the epoch. Same as the ZFR, the ZFF method contains two cascades of ZFR and it will maintain the information and highlight the characteristic of discontinuities around zero frequency. The only difference is that the ZFF also contains one differentiator and two cascade Detrenders. The block diagram of ZFF is shown as
\begin{figure}[htbp]
\centerline{\includegraphics[width=0.5\textwidth]{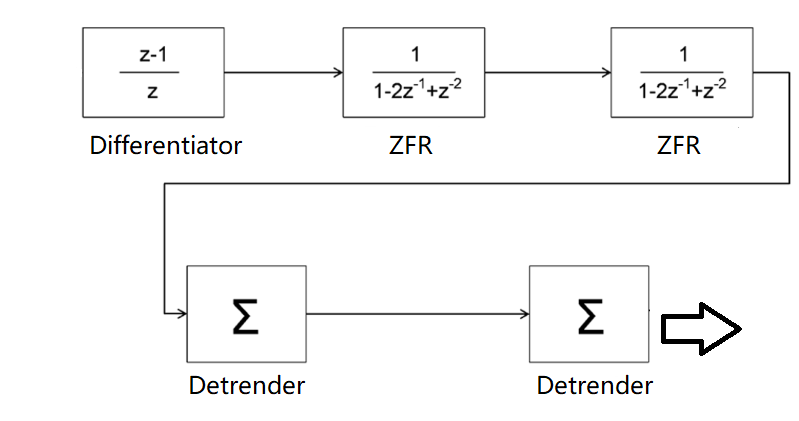}}
\caption{Block Diaram of Zero Frequency Filter}
\label{fig6}
\end{figure}
\begin{itemize}
\item The ZFF method consists of 3 different parts: Differentiator, two cascade ZFR, and two cascade Detrenders. The differentiator is designed to remove the fluctuation at low-frequency components. The equation of differentiator is given as
\begin{equation}
 x[n]=s[n]-s[n-1]\label{eq9}
\end{equation}

\item Pass the output of the differentiator to two cascade zero frequency resonators, the output of the resonator is extremely high because the denominator of the ZFR transfer function is four degrees polynomial so the output signal will grow exponentially. The function of two cascade ZFRs is given as
\begin{equation}
 y_{Two ZFR}(z)=\frac{1}{1-4z^{-1}+6z^{-2}-4z^{-3}+z^{-4}}\label{eq10}
\end{equation}

\item In order to remove the exponential trend in the previous step, the Detrender is applied to get the mean subtraction of the filtered result. After applying the Detrender, the discontinuities will be highlighted because the resulting signal will fluctuate horizontally, not exponentially. The equation of Detrender is given as
\begin{equation}
 y_{Detrender}[n]=y[n]-\frac{1}{2N+1}\sum_{m=-N}^{N}y[n+m]\label{eq11}
\end{equation}
\item The Fig.7 shows the results when passing the 100ms original speech signal to two cascade ZFR and two cascade Detrender consecutively.
\begin{figure}[htbp]
\centerline{\includegraphics[width=0.5\textwidth]{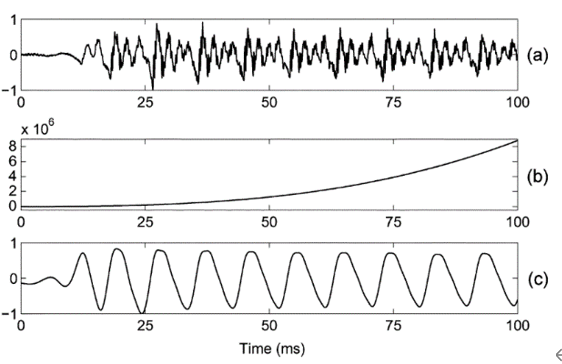}}
\caption{A 100ms of a)Original speech signal b)Pass through  2 cascades Zero Frequency Resonators c)Pass through 2 cascades Detrenders }
\label{fig7}
\end{figure}
\item The equation and frequency response of ZFF is given as
\begin{equation}
H_{ZFF}(z)=\frac{1}{(1-z^{-1})^4}\label{eq12}   
\end{equation}
\begin{equation}
H_{ZFF}(e^{j\omega})=\frac{1}{(1-cos(\omega)+jrsin(\omega))^4}\label{eq13}   
\end{equation}
As shown in Fig.5, the ZFF has a linear phase, which means there is no phase distortion during the extraction of the epochs. This characteristic makes ZFF a robust method to detect and extract epochs with accuracy.
\end{itemize}

\subsection{Propose of Zero Phase Zero Frequency Resonator}
Since ZFF uses Infinite Impulse Response (IIR) filter followed by trend removal blocks however, the filter used in ZFF method is unstable which makes it unsuitable for practical implementation\cite{b6}, another method called Zero-Phase Zero Frequency Resonator (ZP-ZFR) is proposed.
\begin{itemize}
\item The ZP-ZFR consists of 3 parts: pre-emphasis, which is a differentiator, Resonator, and Detrender. The block diagram of ZP-ZFR is given as
\begin{figure}[htbp]
\centerline{\includegraphics[width=0.5\textwidth]{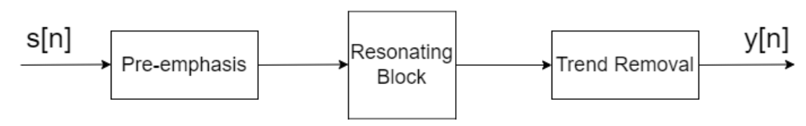}}
\caption{Block diagram for Zero Phase-Zero Frequency Resonator }
\label{fig8}
\end{figure}
\item The function of ZP-ZFR is given as
\begin{subequations}
\begin{align}
H_{ZP-ZFR}(z)&=H(z)H(z^{-1}) \\ 
&=\frac{1}{(1-rz^{-1})^2}\frac{1}{(1-rz)^2} \\
&=\frac{-z^{-2}}{(1-rz^{-1})^2(1-rz)^2} 
\end{align}
\end{subequations}
In Eq.14, the system has poles at $z_1=r$ and $z_2=r^{-1}$, the r is determined by the bandwidth of the system and is regularly ranged from 0 to 1. In order to make sure the system is stable, r should be chosen from 0.95 to 0.99. Generally, the system can be stable by choosing r less than 1.
\item  The Frequency response of ZP-ZFR is given as
\begin{equation}
H_{ZP-ZFR}(e^{j\omega})=\frac{1}{(1-2rcos(\omega)+r^2)^2}\label{eq15}   
\end{equation}
As shown in Fig.5, the ZP-ZFR also has a linear phase, which means there is no phase distortion in the result of epoch detection. Hence, ZF-ZFR can also extract the location of epochs with accuracy and stability.
\end{itemize}

\subsection{Comparison of the methods}
\begin{table}[htbp]
\caption{Comparison of each methods}
\begin{center}
\begin{tabular}{||c c c c||} 
 \hline
 Method & Causality & Phase & Stability \\ [0.5ex] 
 \hline\hline
 ZFR & Causal & Non-linear & Stable \\ 
 \hline
 ZFF & Causal & Linear & Unstable \\
 \hline
 ZP-ZFR & Non-causal & Linear (Zero Phase) & Stable \\
 \hline
\end{tabular}
\end{center}
\end{table}

The Table I above illustrates the three types of filter proposed above and compare them in terms of stability, phase and causality.

For the ZFR method, the pole for that filter is $z=r$ with $0<r<1$, which means all the poles are inside the unit circle, so the ZFR filter is stable. In addition, as the ZFR plot shown in the Fig.5, ZFR is non-linear phase. Besides that, since the Region of Convergence (ROC) can be represented as $z>r$, the system is causal, which means the system is causal.

For the ZFF method, the pole is $z=1$, which means it is on the unit circle. The ROC is $z=1$ and the phase is linear phase so the system is unstable and causal.

For the new proposed ZP-ZFR method, there are two poles locating at $r^{-1}$ and r with $0<r<1$. The ROC is like a ring which covers the unit circle. Hence, the system is stable and non-causal, which means the current output depends on future input. It works better than previous two methods since it will not be influenced by previous information.

\section{IMPLEMENTATION AND EVALUATION}

\subsection{Implementation ZFF and ZPZFR in matlab}
\begin{itemize}
\item As explained earlier, these algorithms consist of a few parts: 1. Pre-emphasize the signal, 2. Apply the filter, 3. Remove Trends in filtered signal, and 4. Keep only the instances of positive crossings (crossing the x-axis from a negative number to a positive number) for ZFF or the negative peaks for ZP-ZFR of the trend removed signal.
\item As for the code specifically, each step will be examined on how to be accomplished. The first step is the easiest to complete simply using the diff() function. Next, the filter is applied using the Eq.10. After this, the signal is passed four times through the trend removal function. This function takes a time interval’s worth of data points and subtracts each data point by the total mean for that window of time. This window of time is typically 15 ms and is iterated so that a window is found that moves throughout the whole signal. This process typically results in an end anomaly which will be removed after the trend removal step. Finally, the epochs can be found isolating the instance of positive crossing.
\item Then the verification should be done to ensure the code is worked. This was accomplished using the Carnegie Mellon University (CMU) Arctic Database. The Arctic database provided an audio signal as well as a complimentary electroglottograph (EGG) signal. This EGG signal measures how much electricity flows across the larynx resulting in a higher voltage the more the vocal cords are touching. For this reason, the negative peaks of the EGG’s derivative can be calculated to find the instances of epochs. In the Fig.9, the epochs which are found by the above method (which is in red) are in the same spot as the EGG’s instances of epochs (the negative peaks in black), thus verifying the extracting process is correct. An amplitude of -1000 to the epochs is assigned for ease of viewing.  
\end{itemize}
\begin{figure}[htbp]
\centerline{\includegraphics[width=0.5\textwidth]{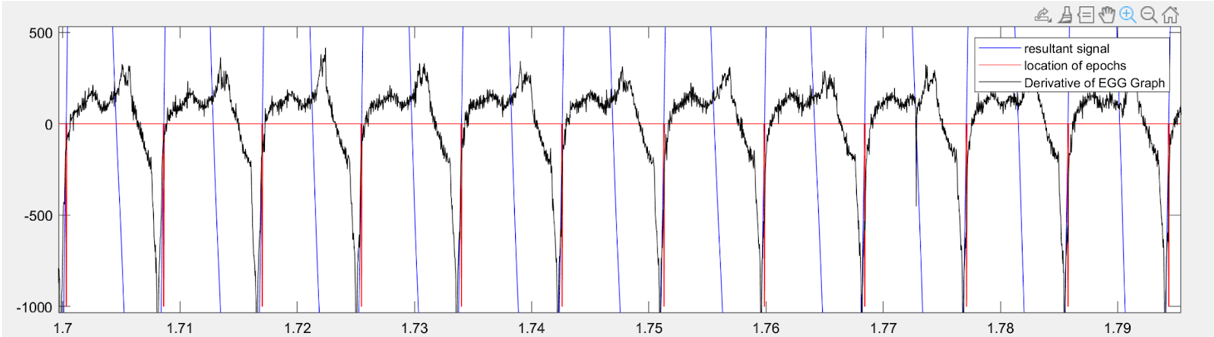}}
\caption{Electroglottograph signal for speech signal }
\label{fig9}
\end{figure}
\subsection{Epoch comparison}
After detecting epochs, the way in which the epochs can be utilized is considered. One of the directions is speech separation. Thus, the epoch in each speech should be compared.  For the epoch comparison function, the epochs’ instances of time will be stored and the Deltas between concurrent epochs will be found and marked (i.e. epoch\_t2-epoch\_t1, epoch\_t3-epoch\_t2, and so on). The previous steps will be performed on both signal 1 and signal 2, but then the time difference between signal 1’s Deltas and signal 2’s Deltas will be computed. From this, one of the methods is to add up how many data points are very close to 0 in the time difference between Deltas. If the number of data points very close to 0 is high, meaning many epochs in one signal are very similar to other signals’ epochs, then it is likely the same speaker. 
\begin{figure}[htbp]
\centerline{\includegraphics[width=0.5\textwidth]{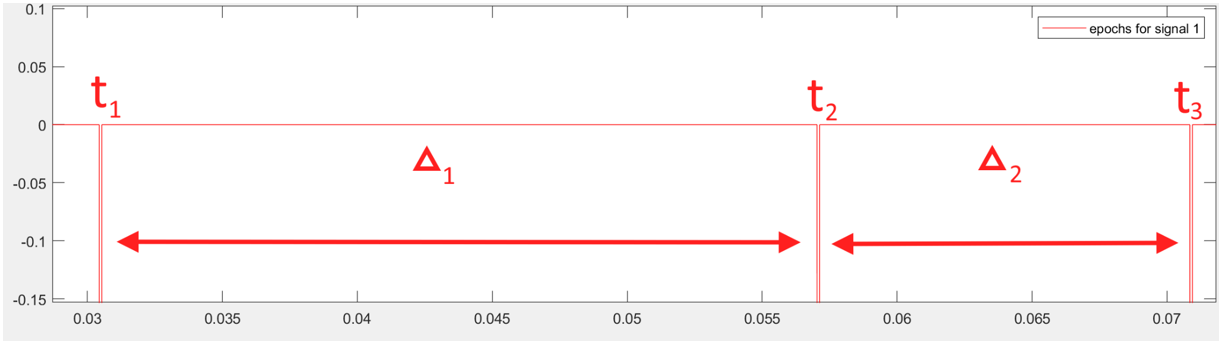}}
\caption{Delta in the speech signal}
\label{fig10}
\end{figure}

\subsection{Speech Differentiation}
After being able to compare epochs, the next step is to use this ability to see if a system could differentiate speakers. To do this, a series of tests with an array of different speakers is set up. Each speaker was requested to say “The University of Mississippi” 7 times. \#1, 2, and 3 would be saying the phrase normally, and \#7 was also said normally which was the “lock” file with which all other files, both from the same speaker and from different speakers, would be compared to. \#4 was saying the phrase fast, \#5 was saying it slowly, and \#6 was saying it in a “wacky” way. This wacky recording was to give the system an anomaly and see how it would respond. As for what “wacky” specifically consisted of was left up to the speaker, an example of one of the wacky recordings was an irish accent. An example of this can be seen in the table II. In this example, each of the speakers' 6 audio recordings were compared to Ferro’s “lock” audio file and the number of Delta\_12 provided a number result which acts as a speaker differentiation confidence number. From this, it is obvious that if it is the same speaker the values are much higher than for the other speakers across the board. While the non-normal cases have a lower value than normal cases, the values are typically higher than for the other speakers. These results can also be seen in the Fig.11. Based on some tests, it can be determined that if the resulting number is above ~7 it is the same speaker. However, the resulting number is never absolute which creates some errors when the system is changed up. For example, when the microphone is switched, it can be found the threshold amount should be changed to ~12 instead of 7. It is also heavily reliant on the specific audio files in question. For these reasons, errors in the system were somewhat common. 
\begin{table}[htbp]
\caption{Number of Delta for different speaker}
\begin{center}
\begin{tabular}{||c c c c c c c||} 
 
 \hline
 Speaker & \multicolumn{6}{|c|}{\textbf{Number of Delta\_12}}\\ [0.5ex] 
 \hline\hline
 Ferro & 7.6 & 8.6 & 12 & 3.3 & 5.3 & 5.6 \\ 
 \hline
 Robinson & 4.3 & 5 & 5.3 & 3 & 1.6 & 2.6 \\
 \hline
 Mcsparin & 2 & 4 & 2.6 & 1.3 & 2.3 & 1.6 \\
 \hline
 Locum & 2.3 & 2 & 2.6 & 2 & 2 & 2 \\
 \hline
 Bradford & 3.6 & 4.6 & 5 & 7 & 4.6 & 1.3 \\
 \hline
\end{tabular}
\end{center}
\end{table}
\begin{figure}[htbp]
\centerline{\includegraphics[width=0.5\textwidth]{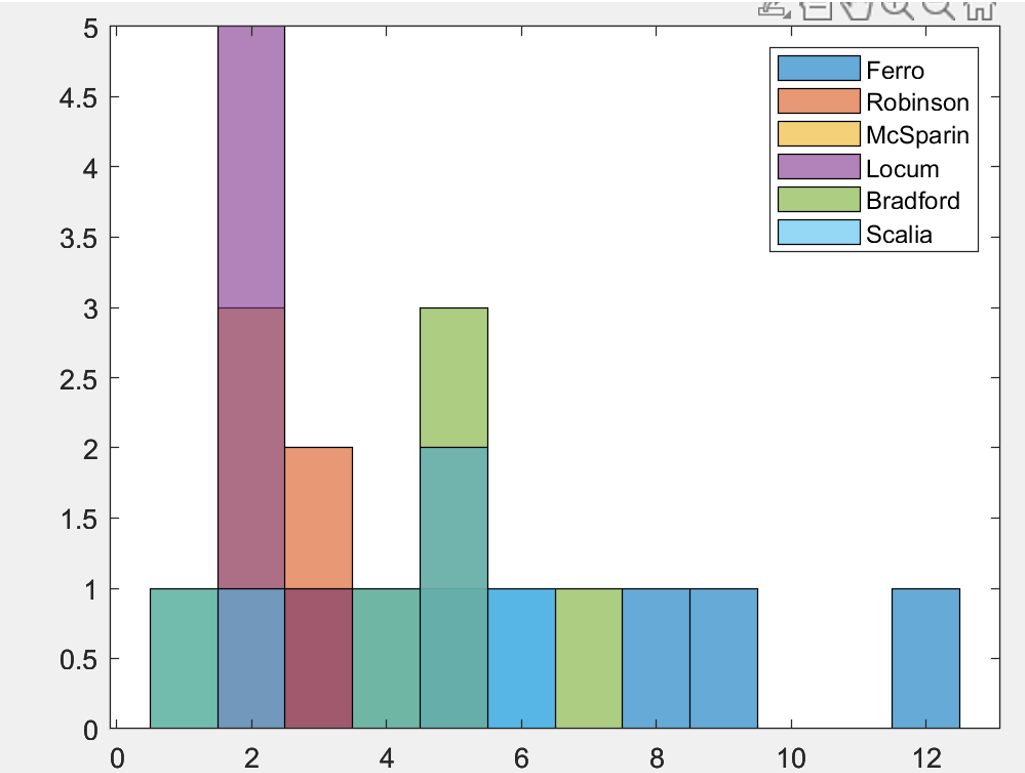}}
\caption{Histogram for the Speaker Differentiation Test}
\label{fig11}
\end{figure}
\subsection{Application case and implementaion}
\begin{itemize}
\item The idea above about speaker identification can be expanded to a prototype of a real-world product. The application was a voice activated lock. This could be used maybe on a door where only the registered speaker would be allowed in or also when a credit card company calls and wants to confirm it is the right speaker. The lock prototype utilizes a raspberry pi, python, and Matlab in order to work properly and the specifics will be covered later. The lock will first record a series of lock files in order to key the lock. Next, a test file will be recorded. If it is the same speaker, the lock will open which is denoted by a green LED and if it is not the same speaker, the lock will stay closed, denoted by a red LED. The system records the audio files on the raspberry pi and then sends the files over to another computer to process the audio files. In Matlab, the epochs of all audio recordings are found and then the test file is compared to all the lock files and averaged. If the resulting average is higher than the threshold, Matlab will send a signal back over to the raspberry pi to open the lock or to keep the lock closed otherwise. Furthermore, raspberry pi and Matlab communicate through a shared file directory and will delete the test file once it is done with it to save on memory. Originally, the system was set-up for one-time use and was improved upon to constantly be running. As stated previously, more specific details will be shared in later parts.  
\item The method requires both Matlab and Raspberry pi implementation. The Raspberry pi records five base files and one test file and sends them to the laptop. The Matlab code is infinitely and continuously running. First, the presence of lock files will be checked. When all five lock files exist, use ZP-ZFR which was mentioned before to compute the epochs in each base file. Then, the presence of the test file will be checked. If the test file also exists, it will calculate the speaker differentiation confidence number and take the average of that number across all of the lock files. Finally, it will compare the average with the threshold. If the average is not 0 and equal or greater than the threshold, indicating that the same person is speaking, Matlab sends a signal to the Raspberry pi to “open” the lock. If the average is smaller than the threshold, which means it is not the same person speaking and sends a signal to “close” the lock. If there is no input detected, the program waits for one second before continuing. The system will also reset if these lock files are deleted.
\end{itemize}

\section{RESULT AND CONCLUSION}
As the implementation illustrated above, the raspberry pi can record the speech by microphone and share the audio file to another computer in which the Matlab is running continuously. Then the lock file will be compared with the test file to show whether the speech is from the same speaker. After receiving the comparison result from Matlab, the Raspberry pi will be forced to open the file “1” or the file “0”. They will trigger the green or red LED through general-purpose input and output (GPIO). The value of Delta can also be treated as a measurement of the accuracy of the epoch detection method. The result of the speaker differentiation by implementing ZFF and ZP-ZFR methods is shown in table III and Table IV respectively.
\begin{table}[htbp]
\caption{Number of Delta for different speaker by implementing ZFF method}
\begin{center}
\begin{tabular}{|c|c c|c|} 
 \hline
 Lock & Test & Number of Delta & Maximum \\
 \hline\hline
  & Ferro & 7.3333 & \\
 \cline{2-3}
  & Ferro & 8.0000 & \\
 \cline{2-3}
  & Ferro & 7.3333 & \\
 \cline{2-3}
Ferro & Li & 5.3333 & 120 \\
 \cline{2-3}
  & Xu & 2.3333 & \\
 \cline{2-3}
 & Syed & 4.0000 & \\
 \hline
 \end{tabular}
\end{center}
\end{table}
\begin{table}[htbp]
\caption{Number of Delta for different speaker by implementing ZP-ZFR method}
\begin{center}
\begin{tabular}{|c|c c|c|c|} 
 \hline
 Lock & Test & Number of Delta & Maximum & increment \\
 \hline\hline
  & Ferro & 14.3333 & & 95.45\%\\
 \cline{2-3}\cline{5-5}
  & Ferro & 16.0000 & & 100.00\%\\
 \cline{2-3}\cline{5-5}
  & Ferro & 21.0000 & & 186.00\%\\
 \cline{2-3}\cline{5-5}
Ferro & Li & 6.3333 & 180& 18.75\% \\
 \cline{2-3}\cline{5-5}
  & Xu & 5.0000 & & 114.29\% \\
 \cline{2-3}\cline{5-5}
 & Syed & 6.3333 & & 58.33\% \\
 \hline
 \end{tabular}
\end{center}
\end{table}

As shown in Table III, the term Maximum means the value of Delta when comparing two identical speeches, in other words, it can indicate the total number of epochs in one episode of speech. The remained part of the table shows the average Delta is about 7.5 for the same speaker and it will decrease to 3.9 for the different speaker. Therefore, by implementing the ZFF method, the system will remain 6\% epochs for the same speaker and roughly 3\% epochs for different speakers.

In Table IV, the total number of epochs in the original speech is 180 and the average of Delta for the same and different speakers is 17.1 and 5.9, respectively, which indicates 10\% epochs will have remained for the same speaker and 3\% for the different speakers. Compared with the ZFF, ZP-ZFR can extract 33.3\% more epochs from the same speech, double the number of Delta found from the speech of the same speaker while slightly increasing the value of Delta from the different speaker. Hence, using ZP-ZFR can more easily distinguish the different speakers from comparing epochs.

By comparison, the ZP-ZFR can provide more accuracy in epoch detection than ZFF method. It can increase the overall epoch detected rate by about 4\%. Besides that, comparing with the ZFR method, the ZP-ZFR still provide more accurate results in epoch detection because it has linear phase. In addition, the characteristic of high stability of ZP-ZFR method assures there is no polynomial growth or decay produced in the results, which is quite common in the ZFF method. However, one of the issues for ZP-ZFR method is the power and the hardware requirement for the ZP-ZFR is much higher if it is implemented on other hardware like FPGA. Hence, the hardware consumption needs to be polished in the future.

\vspace{12pt}

\end{document}